\documentclass[]{elsart}

\usepackage{graphicx}
\usepackage{color}

\usepackage{mathptmx, courier, pifont}
\usepackage[scaled=0.92]{helvet}
\usepackage[T1]{fontenc}
\usepackage{textcomp} 

\newlength{\figdn}
\newcommand{\fplus}[1]{F^+_{#1}}
\newcommand{\fminus}[1]{F^-_{#1}}

\newcommand{\eq}[1]{Eq.~(\ref{#1})}

\newcommand{\fig}[1]{Fig.~\ref{fig:#1}}

\newcommand{\tsub}[1]{_{\mbox{\scriptsize#1}}}

\newcommand{\units}[1]{\mbox{\ #1}}
\newcommand{\isotope}[2]{\mbox{$^{#1}$#2}}

\newcommand{\singlefig}[6]{%
\begin{figure} \vspace{#3}%
\begin{flushright}%
\includegraphics*[scale=#5]{#2}%
\end{flushright}%
\caption{\label{fig:#1} #6}%
\vspace{#4}%
\end{figure}}

\begin{document}

\begin{frontmatter}

\title{Explicit Integration with GPU Acceleration for Large Kinetic Networks}

\author{
Benjamin Brock,$^{1,3}$}
\author{
Andrew Belt,$^{2,3}$}
\author{
Jay Jay Billings,$^{3,5}$}
\author{Mike Guidry$^{2,3,4}$
}

\ead{guidry@utk.edu}

\address{$^1$Department of  Electrical Engineering and Computer Science, 
University of Tennessee, Knoxville, TN 37996-1200, USA}
\address{$^2$Department of Physics and Astronomy, University of Tennessee,
Knoxville, TN 37996-1200, USA}
\address{$^3$Computer Science and Mathematics Division, Oak Ridge National
Laboratory, Oak Ridge, TN 37830, USA}
\address{$^4$Physics Division, Oak Ridge National Laboratory, Oak Ridge, TN
37830, USA}
\address{$^5$The Bredesen Center for Interdisciplinary Research and Graduate
Education, University of Tennessee, Knoxville, TN 37996-1200, USA}

\begin{abstract} 
We demonstrate the first implementation of recently-developed fast explicit 
kinetic integration algorithms on modern graphics processing unit (GPU) 
accelerators. Taking as a generic test case a Type Ia supernova explosion with 
an extremely stiff thermonuclear network having 150 isotopic species and 1604 
reactions coupled to hydrodynamics using operator splitting, we demonstrate the 
capability to solve of order 100 realistic kinetic networks in parallel in the 
same time that standard implicit methods can solve a single such network on a 
CPU. This orders-of-magnitude decrease in compute time for solving systems of 
realistic kinetic networks implies that important coupled, multiphysics problems 
in various scientific and technical fields that were intractable, or could be 
simulated only with highly schematic kinetic networks, are now computationally 
feasible.
\end{abstract}

\begin{keyword}
ordinary differential equations 
\sep 
reaction networks 
\sep 
stiffness
\sep
reactive flows
\sep
nucleosynthesis
\sep
combustion

\vspace{1ex}

\PACS 
02.60.Lj 
\sep
02.30.Jr 
\sep
82.33.Vx 
\sep
47.40 
\sep
26.30.-k 
\sep
95.30.Lz 
\sep
47.70.-n 
\sep
82.20.-w 
\sep
47.70.Pq 

\end{keyword}

\end{frontmatter}

\section{Introduction}

Many important physical processes can be modeled by the coupled evolution of a 
reaction network (kinetic network) and fluid dynamics.  A representative example 
is provided by astrophysical thermonuclear reaction networks, where a proper 
description of the overall problem typically requires multidimensional 
hydrodynamics coupled to the network across a spatial grid.  Within each zone of 
the simulation the hydrodynamical (``hydro'') evolution controls the temperature 
and density, while the network influences the hydrodynamical evolution through 
energy release and composition changes. The solution of large kinetic networks 
by the usual implicit-integration approaches is slow and few calculations have 
attempted to couple the element and energy production strongly to the 
hydrodynamics with a network of realistic complexity.  The most ambitious 
approaches use small schematic networks, perhaps tuned empirically to get 
critical quantities like energy production correct on average, coupled to the 
hydrodynamical simulation.  Then a more physically realistic network is run in a 
separate ``post-processing'' step, where fixed hydrodynamical profiles computed 
in the hydrodynamical simulation with the schematic network are used to specify 
the variation of thermodynamic variables such as temperature and density with 
time.

Many other scientifically-interesting problems employ kinetic networks. 
Representative examples include the networks of chemical reactions required to 
model atmospheric chemistry, chemical evolution networks in contracting 
molecular clouds during star formation, plasma-surface interactions in 
magnetically confined fusion devices, fuel depletion in fission power reactors, 
and chemical burning networks in combustion chemistry. The corresponding 
reaction networks are large. Realistic atmospheric simulations, combustion of 
larger hydrocarbon molecules, studies of soot formation, core-collapse 
supernovae, and thermonuclear supernovae all can involve hundreds to thousands 
of reactive species undergoing thousands to tens of thousands of reaction 
couplings \cite{oran05,hix05}. Current techniques based on implicit numerical 
integration typically are not fast enough to allow coupling of realistic 
reaction networks to the full dynamics of such problems and even the most 
realistic simulations have employed highly schematic reaction networks.

As a representative example, the present situation in Type Ia supernova simulations is
illustrated in \fig{networks}.
\singlefig
{networks}
{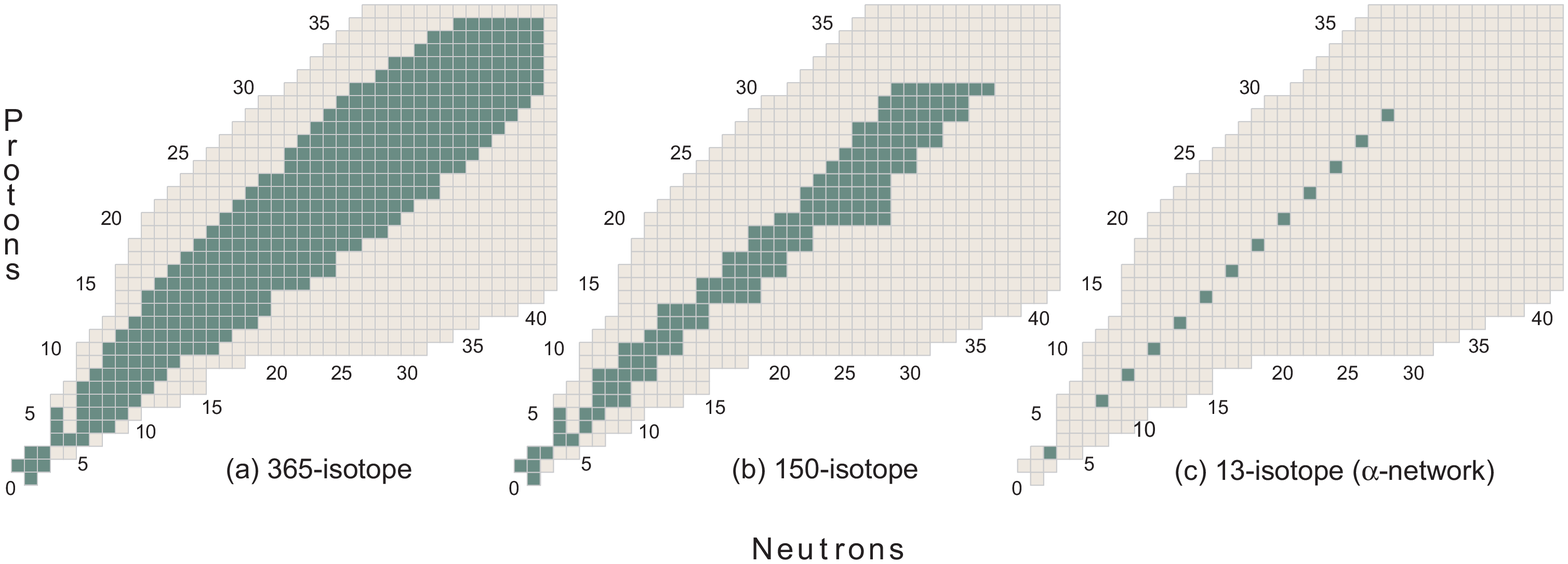}
{0pt}
{\figdn}
{0.46}
{Thermonuclear networks for simulating a typical Type Ia supernova explosion. (a)~A
365-isotope network containing  isotopes that are populated with measurable
intensity in the  explosion.  (b)~A 150-isotope subset of the 365-isotope
network representing the isotopes populated with sufficient intensity to influence
significantly the evolution of the hydrodynamics.  This is the minimal realistic network
for coupling to hydrodynamical simulations of the Type Ia explosion.  (c)~The largest
network (an $\alpha$ network) that has been coupled to multidimensional 
hydrodynamics in a published Type Ia
simulation.}
A physically-realistic network is displayed in \fig{networks}(a), a minimal
physically-correct network for coupling to the hydrodynamical simulation is displayed in 
\fig{networks}(b), and the current state of the art for Type Ia simulations employing
multidimensional hydrodynamics is displayed in \fig{networks}(c). The species omitted in
reducing the 365-isotope network in \fig{networks}(a) to the 150-isotope network in
\fig{networks}(b) are populated sufficiently weakly that they play little role in
energy release and do not have significant influence on the coupling of the network to
hydrodynamics.  Thus, the 150-isotope network is a minimal network for coupling to the
hydrodynamics;  any network smaller than this will omit non-trivial parts of the realistic
coupling between kinetics and fluid dynamics.  The disparity between the minimal realistic
network in \fig{networks}(b) and the current state of the art in \fig{networks}(c) is a
consequence of insufficient computational power to couple a realistic kinetic network in
real time to the fluid dynamics using current technology. This example from astrophysics
is but one example of a number of problems from various fields of science and technology
in which coupling realistic kinetics to fluid dynamics is hampered severely by
insufficient computational speed for kinetic networks. 

There are two general approaches that we might take to address the 
preceding issues. 
 The first is to seek faster algorithms for solution of the typically 
large and stiff system of differential equations that  describe the kinetic 
evolution. The second is to take advantage of advances in computational 
architectures to solve the chosen algorithm more rapidly.  In previous work,\ 
\cite{guidJCP,guidAsy,guidQSS,guidPE}, we described a new 
algebraically-stabilized explicit approach to solving kinetics equations that 
extends earlier work by Mott \cite{mott99} and is capable of taking stable 
integration timesteps comparable to those of standard implicit methods, even for 
extremely stiff systems of equations. Since the methods are explicit, they do 
not involve matrix inversions and thus scale linearly with network size. Because 
of this much more favorable scaling and competitive integration step size, we 
demonstrated that such algorithms are capable of performing numerical solution 
of extremely stiff kinetic networks containing several hundred species 5--10 
times faster than the best implicit codes 
\cite{guidJCP,guidAsy,guidQSS,guidPE}.

One implication of new algorithms is that they may give new perspectives on 
optimization. Standard implicit methods spend most of their time on linear 
algebra operations for larger networks because they must be solved iteratively, 
which requires inversion of large matrices.  Thus, the optimization strategy is 
clear for implicit algorithms: do the linear algebra faster.  Conversely, the 
new explicit methods do not involve matrix inversions, so optimizing them 
involves different strategies.  These may offer unique opportunities for 
implementation on newer architectures such as GPU or many-core accelerators 
coupled to standard CPUs \cite{guidJCP}. In this paper we take a first step in 
addressing these issues by deploying the explicit integration methods described 
in our previous work\ \cite{guidJCP,guidAsy,guidQSS,guidPE} on  coupled
CPU--GPU systems.
We show that using GPUs to exploit the parallelism inherent in the explicit kinetic 
algorithm for networks of realistic size makes it possible find the solution for
a single network on the GPU faster than on the CPU and the solutions for many
networks simultaneously on the GPU versus serially on the CPU.

\section{Implementation of Realistic Kinetic Networks}

We shall assume that the coupling of reaction networks is done using
operator splitting, where the hydrodynamical solver  is evolved for a numerical timestep
holding network parameters constant, and then the network is evolved over the time
corresponding to the hydrodynamical timestep holding the new hydrodynamical variables
constant (see \S\ref{operator-split} below).  The general task for the kinetic 
network then is to solve efficiently $N$
coupled ordinary differential equations 
\begin{eqnarray}
    \frac{dy_i}{dt} &\approx& F_i(y,t) = \sum_j F_{ij}(t) 
\nonumber
\\
&\equiv& \fplus i (t)- \fminus i(t)
    = \fplus i (t) - k_i(t) y_i(t)
\label{eq1.1}
\end{eqnarray}
subject to initial conditions that have been determined in the current
hydrodynamical timestep. In this expression, the $y_i (i=1
\dots N)$ describe the dependent variables (typically measures of abundance),
$t$ is the independent variable (the time in our examples), the fluxes between
species $i$ and $j$ are denoted by $F_{ij}$, and $k_i(t)$ is the effective rate
for all processes depleting the species $i$.  The sum for each variable $i$ is
over all species $j$ coupled to $i$ by a non-zero flux $F_{ij}$, and for later
convenience we have decomposed the flux into a component $\fplus i$ that
increases the abundance of $y_i$ and a component $\fminus i = k_i y_i$ that
depletes it. For an $N$-species network there will be $N$ such equations in the
population variables $y_i$, generally coupled to each other because of the
dependence of the fluxes on the different $y_j$. The variables $y_i$ are typically
proportional to a number density $n_i$
for the species $i$. For the specific astrophysical examples
that follow we shall replace the generic population variables $y_i$ with the
mass fraction $X_i$, which satisfies
\begin{equation}
X_i  = \frac{A_i}{\rho N\tsub A}\, n_i
\qquad
\sum_i X_i =1,
\label{5.35}
\end{equation}
where $N\tsub A$ is Avogadro's number, $\rho$ is the total mass density, and $A_i$ is the
atomic mass number for the species $i$.

\section{\label{algebraic} Algebraic Stabilization of Solutions Using
Asymptotic Approximations}

In the
{\em asymptotic limit} we have $\fplus i \simeq \fminus i$ for the species $i$, leading to
an approximate solution of \eq{eq1.1} given by \cite{guidAsy}
\begin{equation}
y_n = \frac{1}{1+k_n\Delta t} \left(y_{n-1} + F^+_n \Delta t \right).
\label{asySophia}
\end{equation}
Since the asymptotic approximation specified above is expected to be valid if $k\Delta t$
is large, we define a critical value $\kappa$ of $k\Delta t$ ($\kappa = 1$ will be chosen
here) and at each timestep cycle
through all network populations and compute the product $k^i\Delta t$ for each species $i$
and the proposed timestep $\Delta t$.   Then, for each population species $i$
\begin{enumerate}
\item
If $k^i\Delta t < \kappa$, the population is updated numerically by solving \eq{eq1.1}
using a standard explicit forward-Euler algorithm.
\item
Otherwise, for $k\Delta t \ge \kappa$, the population is updated algebraically
using the asymptotic approximation given in  \eq{asySophia}.
\end{enumerate}
This algorithm is explicit, since all quantities required to update a timestep 
are available from the previous timestep.  Thus, it avoids the iterative 
solution and associated matrix inversions required for implicit integration. As 
we have noted above, this algorithm alone implies as much as an order of 
magnitude increase in speed over standard implicit algorithms for solving 
realistic kinetic networks \cite{guidJCP}.  We shall demonstrate below that 
significant additional efficiencies are possible by using the GPU to exploit the 
parallelism inherent in the algebraically-stabilized explicit integration 
algorithm.

\section{\label{operator-split} Operator-Split Integration Timesteps}

In \fig{hydro-networkTimesteps}
\singlefig
{hydro-networkTimesteps}
{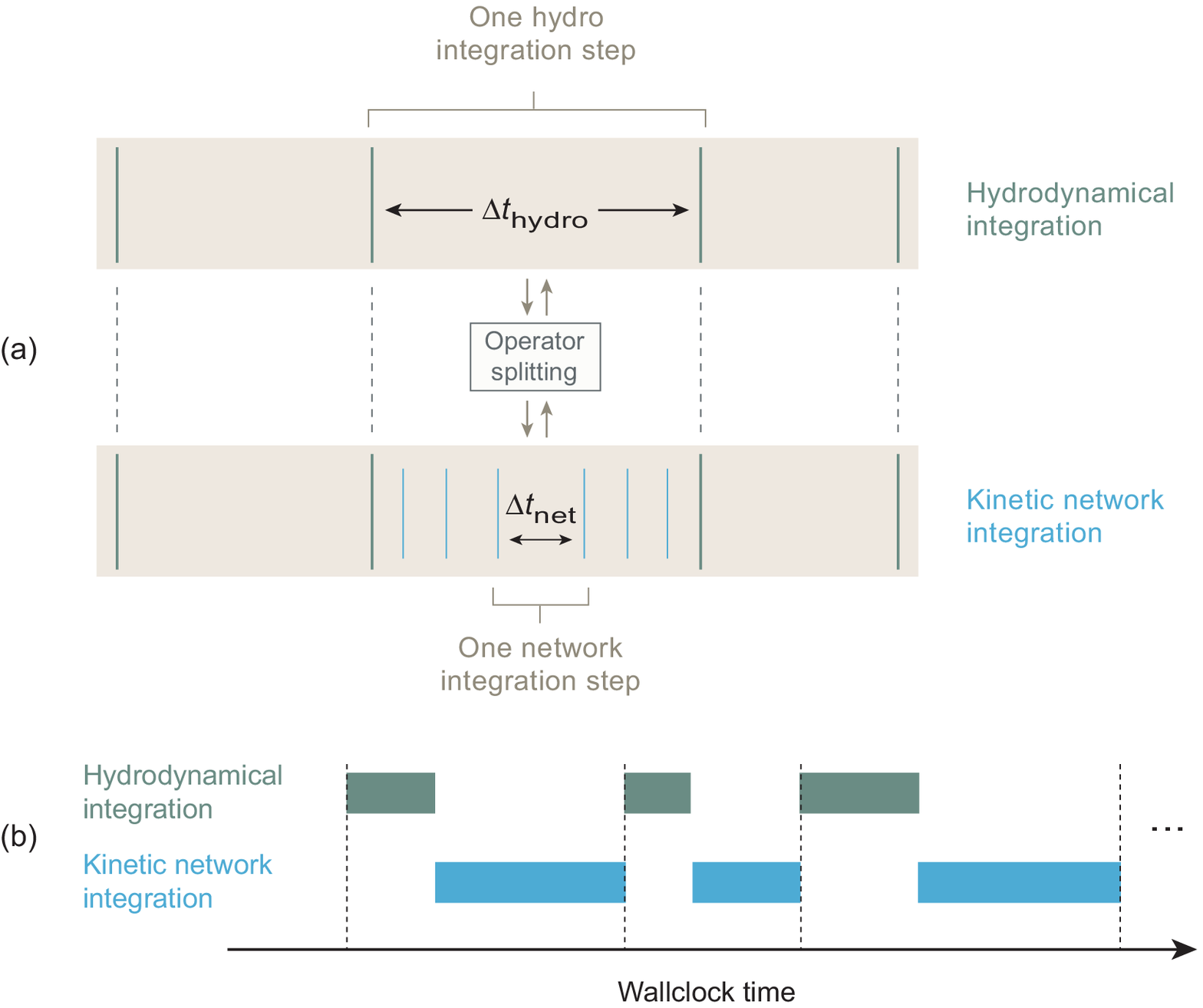}
{0pt}
{\figdn}
{0.72}
{(a)~Illustration of hydrodynamical timesteps $\Delta t\tsub{hydro}$ and kinetic
network timesteps $\Delta t\tsub{net}$ in one hydro zone for operator-split coupling of a
kinetic network to fluid dynamics. The hydro timestep $\Delta t\tsub{hydro}$ sets the
interval over which the kinetic network will be integrated using adaptive timesteps
$\Delta t\tsub{net}$. (b)~The elapsed wallclock time as the operator-split 
integration proceeds, alternating between the hydrodynamical and kinetic 
network integration. One full integration step (hydro plus kinetic network) 
requires the elapsed time between two dashed lines.  Generally, the wallclock 
time to integrate the hydro within a full step is different from that required 
to integrate the network in the same full step.}
the relationship of a hydrodynamical integration timestep $\Delta t\tsub{hydro}$ 
(``hydro timestep'') to a kinetic network integration steps $\Delta 
t\tsub{net}$ (``network timestep'') is illustrated for an operator-split 
simulation. The hydro integrator takes an adaptive timestep $\Delta 
t\tsub{hydro}$ while the kinetic network is dormant. Then the updated 
hydrodynamical variables (temperature, density,  \ldots) are held constant while 
the kinetic network is integrated over the interval $\Delta t\tsub{hydro}$ using 
adaptive network timesteps $\Delta t\tsub{net}$. The updated abundance variables 
and the energy released by the kinetic network are then passed from the kinetic 
network to the hydro integrator, which uses these and the equation of state to 
set the initial conditions for the next hydro integration timestep, and so on. 

The hydro timestep $\Delta t\tsub{hydro}$ and the network timestep
$\Delta t\tsub{net}$ are different timescales and should not be confused in the 
following discussion.  The hydro timestep is set by characteristic times for 
response of the fluid while the network timestep is set by the inverse of the 
reaction rates in the kinetic network.  For the examples discussed here, 
typically $\Delta t\tsub{hydro} \ge \Delta t\tsub{net}$ and the kinetic 
integrator might take $\sim$ 1--1000 network timesteps $\Delta t\tsub{net}$ 
over the interval of one hydro timestep $\Delta t\tsub{hydro}$, depending on 
the ratio of characteristic kinetic reaction times to fluid response times in 
a given zone of the simulation. For our purposes, the hydro integration 
may be viewed as a black box to which the kinetic network is coupled by two-way 
transfer of information, and the only role of the current hydro timescale is to 
set the interval $\Delta t\tsub{hydro}$ over which the kinetic network is to be 
integrated using adaptive timesteps $\Delta t\tsub{net}$.

\section{\label{GPUacceleration} GPU Acceleration}

Modern supercomputers (as well as desktop and laptop systems) often have access 
to GPUs that can greatly accelerate the execution of algorithms formulated to 
take advantage of the parallelism exposed by the GPU. For example, Oak Ridge 
National Laboratory's Titan supercomputer, which has been benchmarked at 17.59 
petaflops and has a theoretical peak performance of 27.1 petaflops, employs 
18,688 compute nodes, each consisting of 16 CPU cores and one NVIDIA Tesla K20X 
GPU, which uses the Kepler GPU microarchitecture \cite{top500}.  The most 
powerful applications must integrate CPUs (each executing a few heavyweight 
threads) and GPUs (each executing many lightweight threads) seamlessly to reach 
speeds that are a significant fraction of the peak capability of the machine.

NVIDIA's CUDA framework provides a parallel computing platform in which 
computational kernels written in CUDA C/C++ can be offloaded to the GPU from 
code running on the CPU. The CUDA programming model utilizes a heterogeneous 
memory paradigm in which a user first copies data from the CPU to the GPU using 
a special CUDA function, then launches a large number of threads, organized in 
blocks, which perform some computation.  Thread blocks are distributed among 
streaming multiprocessors, which contain physical execution cores for integer 
and floating point operations along with registers, thread schedulers, and a 
cache, part of which can be accessed directly by the user as shared memory.  
Once kernel execution is complete, the user may copy data back to the CPU.

GPUs have various limitations that must be overcome 
to achieve large speedup from the massively-parallel, lightweight-thread 
implementation. Four issues are of particular importance:

\begin{enumerate}
 \item 
Data transfer between CPU and GPU is slow relative to compute speeds;
thus scalable CPU--GPU computation must control the cost of communication
between the CPU and GPU.
\item
Code running on GPUs must be highly parallel,
since even small serial portions of a parallel code will greatly diminish
performance (Amdahl's Law).
\item
A modern general-purpose GPU typically has large amounts of relatively slow
global memory and smaller amounts of much faster shared memory. For example, on a
Tesla K20X GPU, each thread block has access to 6 GB of slow global memory and 48
KB of fast memory shared only within the block. Thus, optimal use of the GPU
must enable a significant fraction of the calculation to use the fast, shared
memory.
\item
The highest amount of performance is gained when as many as
possible of the total number of threads on the GPU are used (``occupancy'').
\end{enumerate}
Because of the simplicity and associated transparency of explicit integration 
algorithms and their intrinsically parallel nature for many of the required 
computing tasks, they are particularly attractive candidates for parallelization 
using GPUs. This task is greatly facilitated by the availability of computing 
platforms that allow high-level access to the GPU without 
the user having to deal with the details of thread management.  In the 
applications discussed in this paper we have used NVIDIA CUDA C/C++ to implement 
algorithms that are launched under CPU control but that execute entirely on the 
GPU.

It is important to note that we do not present a detailed discussion on what the
CPU can do with its free cycles. Ideally, the CPU would solve other coupled
physics problems with these cycles in such a way that maximizes the amount of
work done but minimizes the amount of time spent waiting for either piece of
hardware to finish. However, it is also possible that the CPU could work on
post-processing or uncertainty quantification. The easiest way to consume the
cycles is to launch the GPU work asynchronously on a separate thread, possibly
using OpenMP, and then do the CPU work.

\section{Implementing the Explicit Asymptotic Algorithm on a CPU--GPU 
System}

The prototype implementation that we shall describe here assumes an 
operator-split formulation of fluid dynamics coupled to a large kinetic network, 
with the computation of the fluid dynamics implemented on the CPUs and the 
computation of the kinetic networks implemented on the GPUs. This framework 
describes qualitatively a large number of potential scientific applications in a 
variety of fields, but to be definite we shall emphasize  astrophysical 
thermonuclear networks coupled to hydrodynamical simulations in explosive 
burning scenarios. Our reference example will correspond to a 150-isotope 
network containing 1604 reactions (the minimal realistic network of 
\fig{networks}(b)),  integrated at a constant temperature of $7 \times 10^9$ K 
and constant density of $10^8 \units{g\,cm}^{-3}$, but we shall display 
calculations with as many as 365 network species and 4300 reactions.  In 
realistic simulations one often encounters approach to equilibrium and must use 
the asymptotic algorithm described above supplemented by a partial equilibrium 
approximation \cite{guidJCP,guidPE}. Adding the partial equilibrium algorithm 
should not affect the parallelism of the problem substantially, so we shall 
simplify and illustrate using calculations not near equilibrium where the 
asymptotic approximation is sufficient.

In \fig{CUDAnetworkFlow}
\singlefig
{CUDAnetworkFlow}
{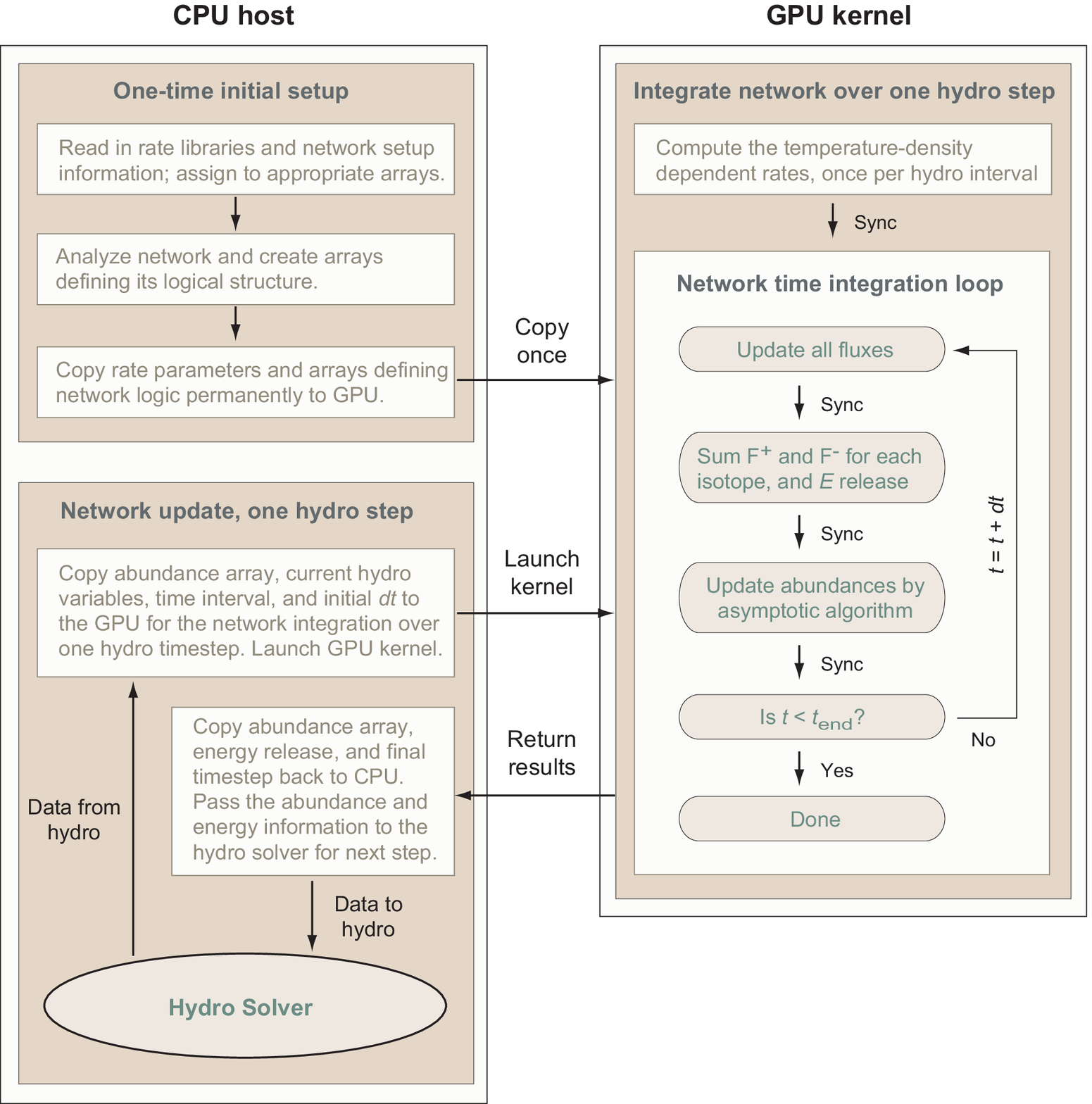}
{0pt}
{\figdn}
{0.65}
{Schematic flow for the example calculation presented here for a single kinetic 
network. The kinetic network is assumed to be coupled to the hydrodynamical 
integration by operator splitting. The kinetic network integration over a time 
interval corresponding to one hydro timestep is executed entirely on the GPU.  
Once the problem is set up, the only communication between CPU and GPU is to 
copy data from the last hydro timestep to the GPU, launch the kernel, and then 
copy the network integration results back to the CPU. The points labeled 
``Sync'' in the GPU kernel are points where the algorithm requires that all 
threads be synchronized before proceeding because subsequent operations require 
the completed results of those threads.}
we illustrate the basic structure of our example calculation. We shall refer to 
the code running on the CPU as the {\em host} and the code running on the GPU as 
the {\em kernel} in the following discussion.  In a setup step that is executed 
once for each overall problem, library parameters required to calculate the 
temperature and density-dependent reaction rates are copied to the GPU and are 
resident on the GPU for the duration of the calculation. At the end of each 
hydrodynamical integration step a fully coupled CPU code would hold initial 
conditions for the kinetic network, which consists of the current temperature 
and density, and the current abundances for all species in the kinetic network. 
In a realistic simulation these values would be supplied to the CPU code by the 
hydrodynamical integration, but the source is irrelevant for our present tests 
and in our simulation we simply read in a trial set of initial conditions for a 
hydro timestep. One full kinetic network integration over a time interval 
corresponding to one hydrodynamical timestep then consists of the following 
steps.
\begin{enumerate}
 \item
Copy from the CPU to the GPU (1)~a vector of current network abundances, 
(2)~the temperature, density, and duration of the current hydro timestep 
$\Delta t\tsub{hydro}$, and (3)~a trial initial network timestep. For the 
representative 150-isotope network this corresponds to copying a total of 154 
floating point numbers from the CPU to the GPU.
\item
Launch a GPU kernel to integrate the kinetic network over the time interval
corresponding to the hydro timestep.
\item
When the integration on the GPU is complete, copy back to the CPU values of 
(1)~the updated species abundances at the end of the kinetic network 
integration, (2)~the integrated energy release over the kinetic network 
integration, and (3)~a final kinetic network integration timestep for use in 
setting the initial trial timestep in the next network integration. For the 
representative 150-isotope network this corresponds to copying a total of 152 
floating point numbers from the GPU to the CPU.
\end{enumerate}
In this scheme, during the kinetic network integration corresponding to one hydrodynamical
timestep (see \fig{hydro-networkTimesteps}) the network integration is done {\em entirely
on the GPU} and the only communication between the CPU and GPU is at the beginning and end
of the network integration. For the examples discussed here, the required data transfer
between the CPU and GPU over each hydro integration step is thus $\sim$1 kB at the
beginning and a similar amount at the end of each network integration. For typical
installations the CPU--GPU data transfer rate is $\sim 10^{11}$ bytes per
second, so each network integration over the interval of a hydro timestep requires a 
communication time that should be a negligible fraction of the total network
integration time.

\section{Performance for a Single Network}

The accuracy of the GPU calculation relative to the reference CPU implementation 
of the algorithm used in Refs.\ \cite{guidJCP,guidAsy,guidQSS,guidPE} has been 
tested for various networks containing from 14 to 365 isotopic species,
\textit{in both single and double precision.}  This is illustrated in 
\fig{massFractionGPU_CPU_150}%
\singlefig
{massFractionGPU_CPU_150}
{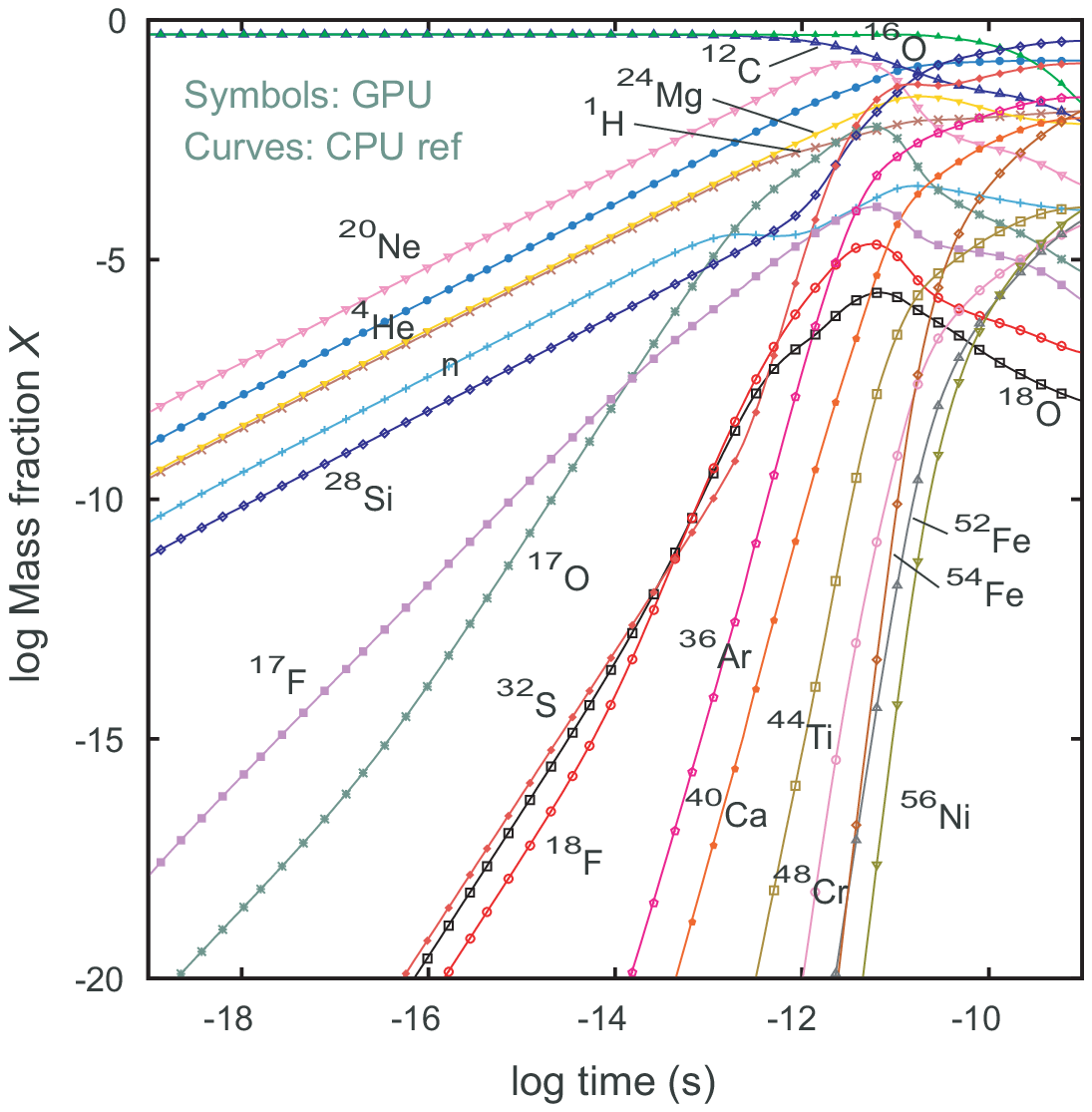}
{0pt}
{\figdn}
{0.80}
{Comparison of selected isotopic mass fractions for 150-isotope calculation 
using the GPU parallel (symbols) and the reference CPU serial (curves) 
implementation of the algorithm, both integrated in double precision. A constant 
temperature of $7 \times 10^9$ K and a constant density of $10^8 
\units{g\,cm}^{-3}$ (conditions typical of a strongly-burning zone in a Type Ia 
supernova simulation) were assumed. }
for some arbitrarily selected species from the representative 150-isotope 
network integrated with double precision. It is clear that in double precision 
the GPU integration gives essentially the same results (typical differences 
between points and curves are less than one part in $10^4$) as the reference CPU 
implementation of the algorithm for mass fractions ranging over 20 orders of 
magnitude.

For GPU applications the limited amount of fast shared memory per block would 
make the use of single rather than double precision in the kinetic network 
highly advantageous, if it leads to stable and accurate results.  We have tested 
single versus double precision implementations of the GPU algorithm.  In general 
we find that for the conditions used in our tests (which are probably as 
extreme as for any kinetics simulation) the single-precision results are 
stable and more than accurate enough for coupling to fluid dynamics simulations. 
 An example for the representative 150-isotope network is displayed in 
\fig{singleVsDouble_150}.%
\singlefig
{singleVsDouble_150}
{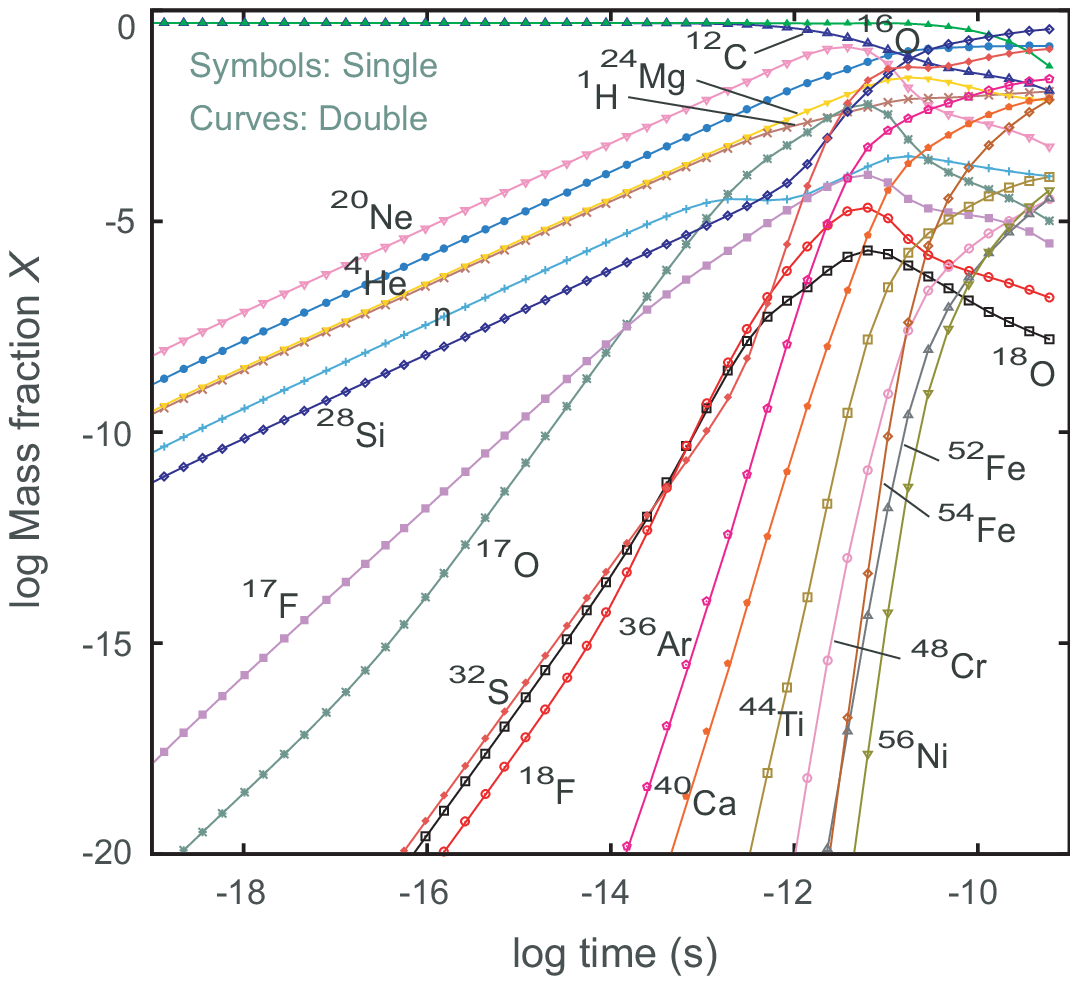}
{0pt}
{\figdn}
{0.80}
{Comparison of GPU calculation of mass fractions in double precision (curves) 
and single precision (symbols). Arbitrarily selected isotopes from a 
150-isotope calculation assuming a constant 
temperature of $7 \times 10^9$ K and a constant density of $10^8 
\units{g\,cm}^{-3}$. }
Thus we anticipate that considerable speed increase may be possible in a 
variety of applications by using single rather than double precision variables 
for the network, which allows more of the calculation to fit in the fast shared 
memory.

The execution time for the explicit asymptotic algorithm on several 
representative GPU installations for the case shown in \fig{singleVsDouble_150} 
is displayed in Table \ref{table:netIntegrationTimes}.%
\begin{table}[t]
  \newcommand{\heavyrule}{\hline}
  \centering
  \caption{150-isotope network integration times on 
various GPUs}
  \label{table:netIntegrationTimes}
  \begin{small}
    \begin{centering}
      \setlength{\tabcolsep}{4 pt}
      \begin{tabular}{cccccc}
        \heavyrule
            System &
            Microarchitecture &
            Time (s) &
            Steps &
            Time/Step (s) &
            Time/100 Steps (s)

        \\        \heavyrule
            Tesla M2090  &   
            Fermi &
            3.581 &
            32,182 &
            $1.11 \times 10^{-4}$ &
            $0.011$

        \\        
            Tesla K20X &          
            Kepler &
            5.910 &
            32,182 &
            $1.84 \times 10^{-4}$ &
            $0.018$
         \\
            GT 640 &         
            Kepler  &
            4.590   &
            32,182   &
            $1.43 \times 10^{-4}$  &
            $0.014$
        \\        \hline
      \end{tabular}
    \end{centering}
  \end{small}
\end{table}
To put things on a common footing, we have divided the total integration time by 
the total number of integration steps and reported the average time to execute 
one network integration step in column 5.  There is a range of almost two in 
speeds for different GPU microarchitectures, but we see that generally a 
GPU implementation is able to execute a kinetic integration step in $0.1$-$0.2$ 
ms.

We may compare these results with representative explicit and implicit serial 
implementations of kinetic network integration on CPUs.  For the example used in 
Table \ref{table:netIntegrationTimes}, the serial implementation of the explicit 
asymptotic  algorithm \cite{guidAsy} took $1.9 \times 10^{-4}$ seconds per 
kinetic integration step on a $\sim$ 3 GHz processor, which is not much longer 
than the GPU speeds displayed in Table \ref{table:netIntegrationTimes}.  
Although the GPU version is more parallel, the processors on the GPU are as much 
as 3--4 times slower than the CPU processor used, which partially offsets the 
parallelism advantage at the present level of optimization for the GPU code. We 
anticipate that with further optimization the explicit GPU code will become 
substantially faster than the corresponding CPU explicit code on present 
architectures.

For comparison with standard implicit methods we use as reference current  
implementations of the backward-Euler implicit code Xnet \cite{raphcode}, which 
is a standard computational tool for solving thermonuclear networks in 
astrophysics and was the implicit-code reference used for comparisons in Refs.\ 
\cite{guidJCP,guidAsy,guidQSS,guidPE}. We shall leave systematic comparisons 
aside until the explicit GPU code is more completely optimized, but some 
immediate quantitative comparisons are possible. We noted above that the 
current explicit serial CPU code is almost as fast as the GPU code at its 
current optimization.  In Ref.\ \cite{guidAsy} we used scaling arguments to 
predict that the serial (algebraically-stabilized) explicit algorithm should be 
able to integrate a 150-isotope network 5 or more times faster than an implicit 
algorithm on the same architecture because of faster computation of each 
timestep.  Currently the implicit code using a sparse-matrix solver requires 
0.5--1.0 ms per integration step on a CPU for integration of the representative 
150-isotope network, depending on whether one or two Newton--Raphson iterations 
are required in an integration step (use of non-sparse methods for a 150-isotope 
network would be $\sim 2$ times slower) \cite{harrisPrivate}.     Thus, with the 
fastest GPUs in Table \ref{table:netIntegrationTimes} the 150-isotope network 
appears to be executing 5--10 times faster on a GPU than current implicit codes 
running on a CPU.   

The last column in Table \ref{table:netIntegrationTimes} displays the time to 
execute 100 network integration steps using the explicit asymptotic GPU solver, 
which would be a representative number of kinetic network timesteps required in 
one hydro timestep for problems of the kind discussed here.
Since the total time of this integration is $\sim 10$ ms, this suggests that a
realistic kinetic network coupled to another physics solver could be
executed in a time that will not slow the integration of the other physical
system to a debilitating degree.

\section{Scaling with Network Size for a Single Network}

 \fig{scaling_isotopes2}%
\singlefig
{scaling_isotopes2}
{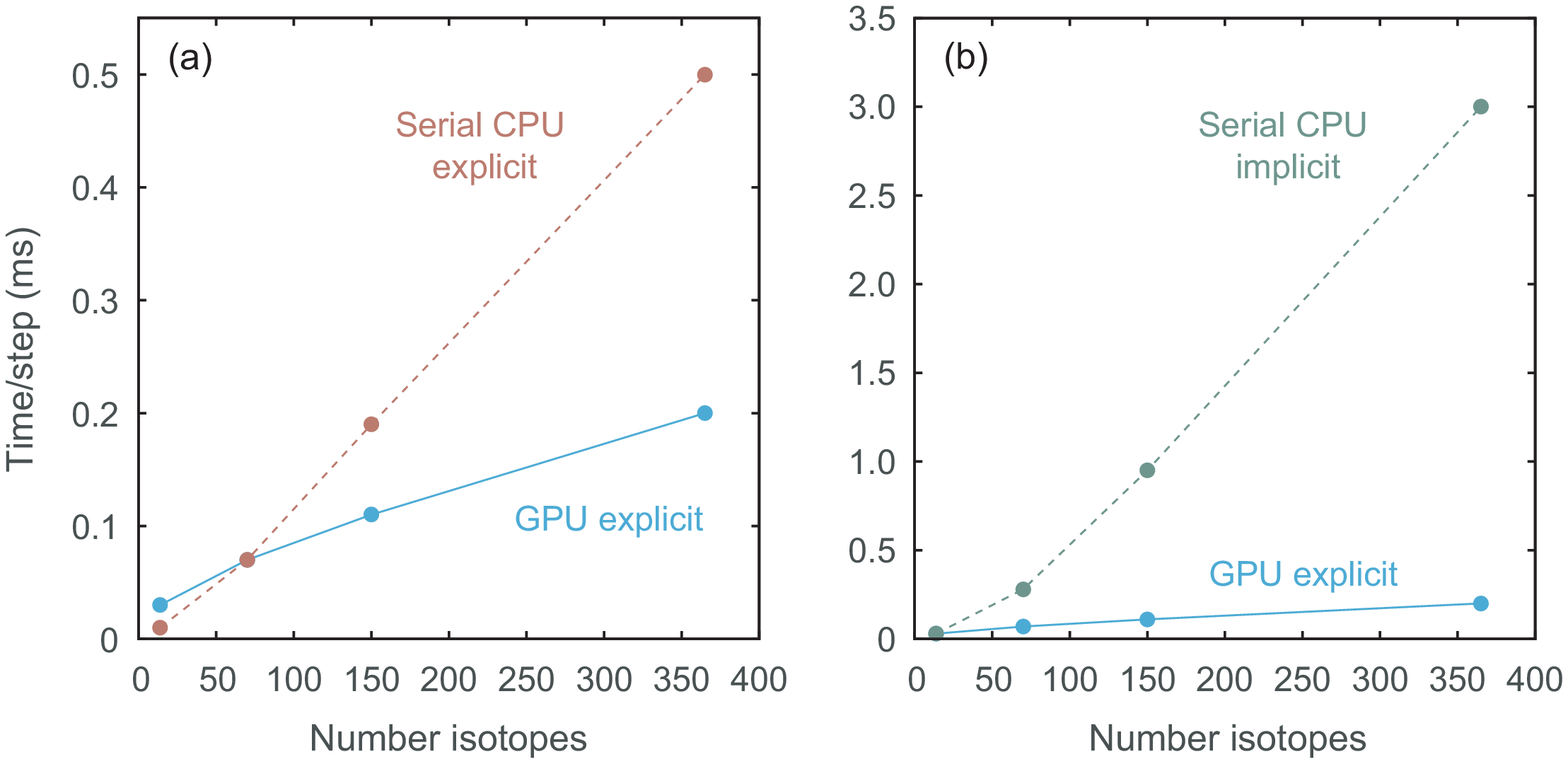}
{0pt}
{\figdn}
{0.60}
{Time to execute a single integration step in milliseconds for one network as a 
function of the number of isotopes in the network.   (a) GPU implementation of 
the explicit asymptotic algorithm (solid blue curve) versus CPU serial 
implementation of the same algorithm (dashed red curve). (b) GPU implementation 
of the explicit asymptotic algorithm (solid blue curve) versus serial 
implementation of a standard backward Euler implicit method (dashed green 
curve). The implicit curve was estimated using the scaling factors $F$ 
determined in Ref.\ \cite{guidAsy} to scale the explicit serial CPU curve in 
(a). See the text for further explanation. All calculations assumed Type Ia 
supernova conditions with a constant temperature of $7\times 10^9$ K and a 
constant density of $10^8 \units{g\,cm}^{-3}$.  GPU calculations were run on 
a Tesla M2090 Fermi architecture; CPU calculations were run on a 3 GHz Intel 
processor.}
shows the scaling of execution time for a single network integration step 
with network size.  In \fig{scaling_isotopes2}(a) we compare the GPU version of 
the explicit asymptotic algorithm with a serial version of the same algorithm 
run on a standard 3 GHz CPU.  We see that the speeds are comparable for small 
networks (with the serial code somewhat faster) but the parallel GPU version is 
approximately twice as fast as the serial version for 150 isotopes and almost 
three times as fast for 365 isotopes.  Although the GPU version is aided by 
added parallelism, this is partially offset by the greater speed of the CPU 
processor versus GPU processors.  

In \fig{scaling_isotopes2}(b) we make an approximate comparison of the parallel 
GPU explicit asymptotic algorithm with a standard backward Euler implicit 
integrator as a function of network size. We obtained the implicit curve by 
multiplying the serial explicit CPU curve in \fig{scaling_isotopes2}(a) by the 
scaling factors $F$  computed as a function of network size in Ref.\ 
\cite{guidAsy}, which give the ratio of the times to compute a single step for 
implicit and explicit methods, with $F > 1$ because of the added matrix 
overhead of the implicit algorithm.  The scaling of this curve could vary by 
factors of several because of variables such as the numerical solver used for 
the implicit method and the relative number of timesteps required in the 
implicit and explicit solves; however, with the present assumptions we see that 
for small networks the speeds are comparable but the explicit GPU calculation 
becomes considerably faster than the implicit calculation as network size 
increases.  For 150 isotopes it is $\sim 9$ times faster, and for 365 isotopes 
the GPU code is $\sim 15$ times faster than the serial implicit calculation. We 
note that the explicit GPU scaling advantage of approximately 9 for a 
150-isotope network inferred from \fig{scaling_isotopes2}(b) is consistent with 
the estimate made above based on current calculations with the implicit backward 
Euler code Xnet, giving some confidence in the validity of the comparison in 
\fig{scaling_isotopes2}(b).  

Comparison of \fig{scaling_isotopes2}(a) and 
\fig{scaling_isotopes2}(b) indicates that the significant speed advantage of 
the GPU code for larger networks has two sources:  (a)~the explicit algorithm 
itself is faster per timestep for larger networks than implicit algorithms, as 
documented in Refs.\ \cite{guidAsy,guidQSS,guidPE,guidJCP}, and (b)~the GPU 
implementation of the explicit algorithm is faster than the serial 
implementation of the same algorithm because of the enhanced parallelism of the 
GPU code, which is sufficient even at the present early stage of optimization 
to outweigh the slower speed of the GPU processors.

\section{Multiple Parallel Networks and Thread Occupancy}

The preceding examples indicate that a single realistic kinetic network can be 
integrated entirely on the GPU, with minimal communication  overhead 
since CPU--GPU transfer of only a small amount of data at the beginning and end 
of the integration is required.  This is already a substantial advance, since 
this implies that fast realistic fluid dynamics coupled to kinetic network 
simulations are now possible with the kinetic network running entirely on the 
GPU, freeing almost all CPU cycles for implementing the fluid dynamics.  
However, implementing a single network on the GPU is a woefully poor utilization 
of available threads since it basically engages only one of the available 
streaming multiprocessors. It is not easy to increase GPU thread occupancy in 
the solution of a single network because the network solution requires 
synchronization at several places (see \fig{CUDAnetworkFlow}).  Without 
returning to the CPU, this can be enforced only within a single block containing 
a maximum of 1024 threads using present technology. However, it is desirable to 
run more than one network at a time on the GPU because in typical applications 
the CPUs of a compute node having a GPU will host multiple fluid dynamics zones 
and each zone  has an independent network reflecting the conditions in that 
zone. Thus, we have tested running many networks in parallel on a single GPU, as 
illustrated schematically in \fig{multipleNetworks}.%
\singlefig
{multipleNetworks}
{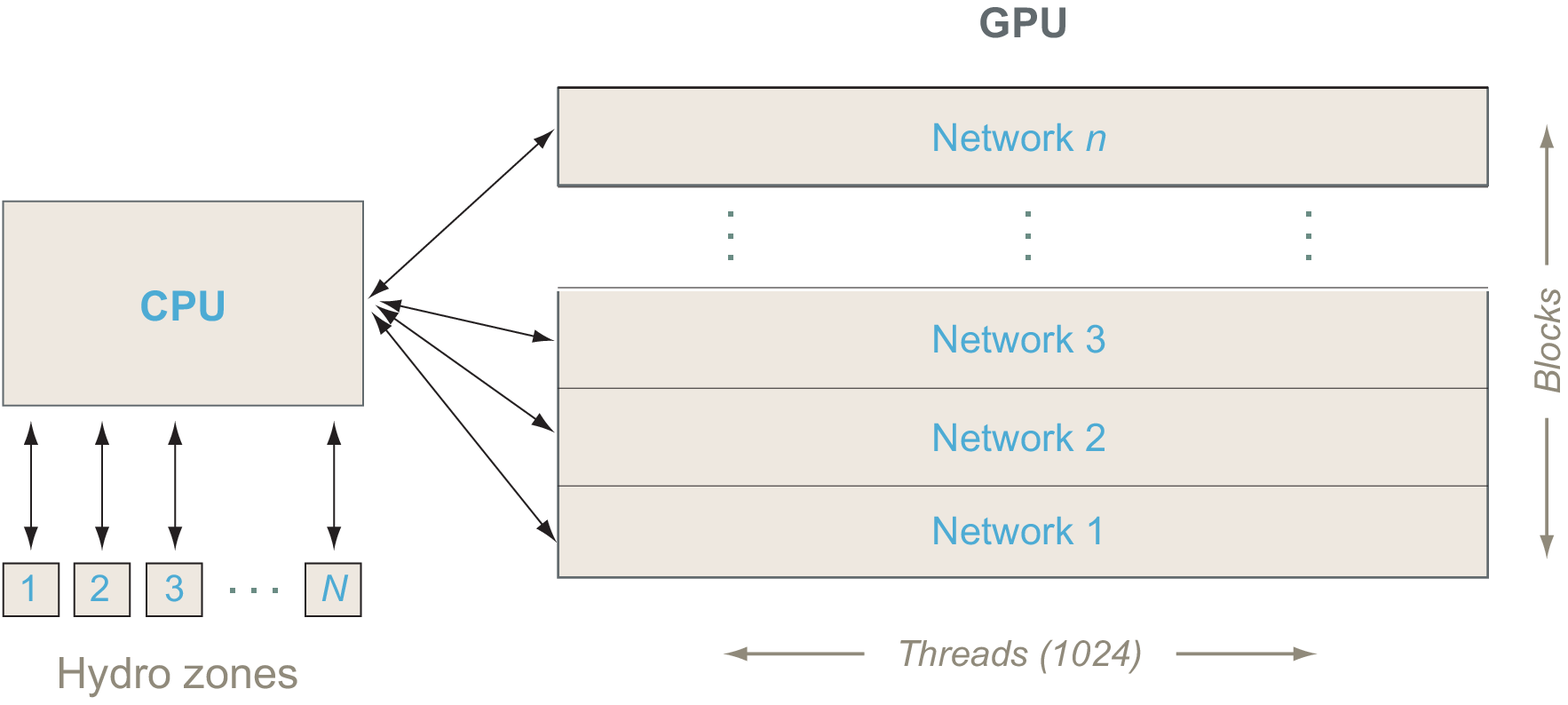}
{0pt}
{\figdn}
{0.60}
{Schematic illustration of integrating multiple kinetic networks in parallel on the
GPU by stacking one network per block.}

In our tests of concurrent execution of the reaction network kernel we deploy a 
number of OpenMP threads, with one asynchronous kernel launched by each thread 
after a small amount of processing  on the CPU to prepare data for 
copying to the GPU.  Timings are taken using the CUDA events timer and are 
started before thread creation and ended after all threads have joined. Thus the 
timing includes CPU processing and copying overhead as well as kernel execution 
time. Timing results for the launch of many representative 150-isotopes networks 
running in parallel are displayed in \fig{stackedNetworkTiming}.%
\singlefig
{stackedNetworkTiming}
{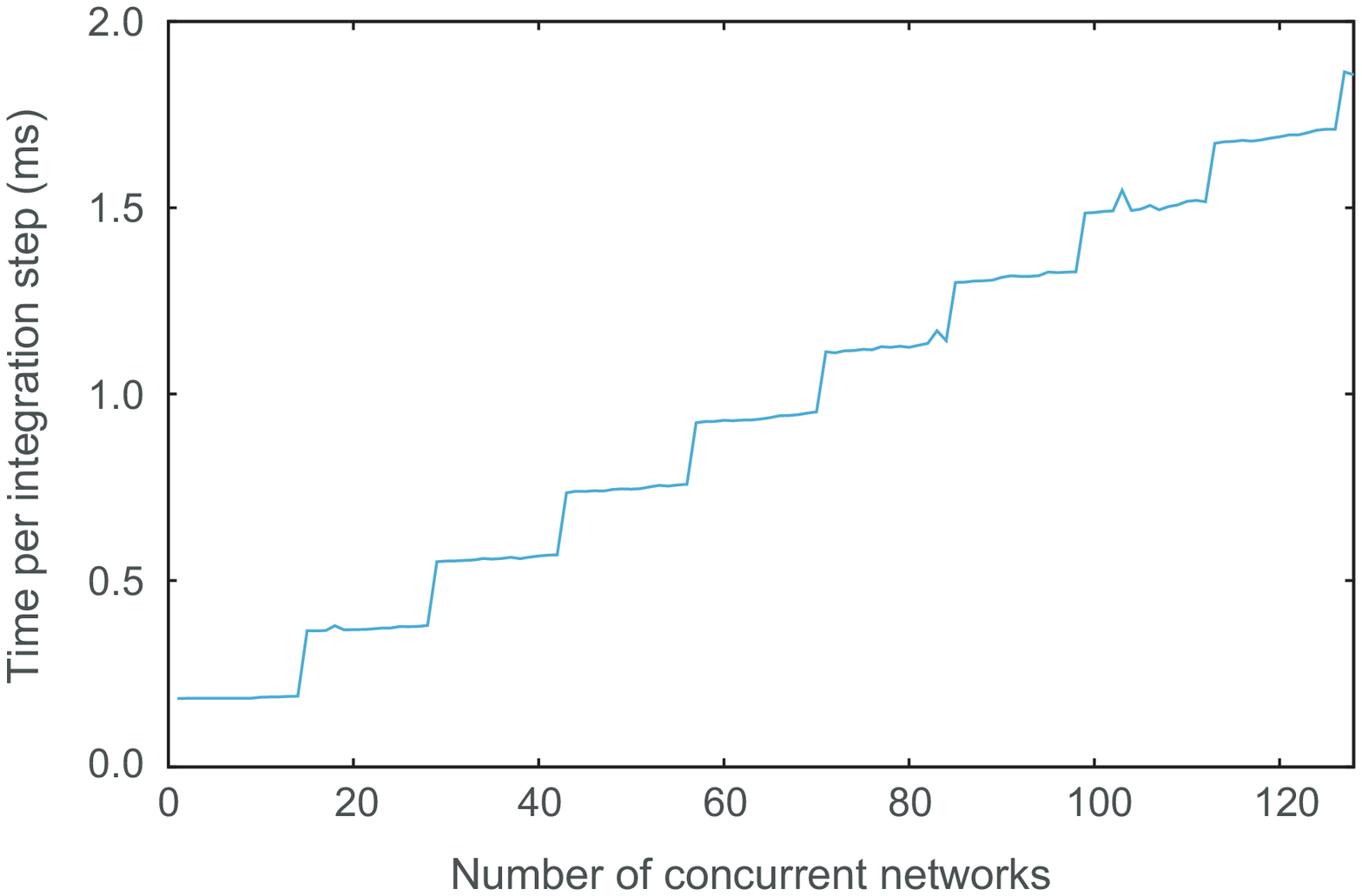}
{0pt}
{\figdn}
{0.55}
{Timing for multiple networks running in parallel on the GPU. All calculations 
assumed Type Ia supernova conditions with a constant temperature of $7\times 
10^9$ K and a constant density of $10^8 \units{g\, cm}^{-3}$.  These 
calculations were run on a Kepler GPU on Titan, which has 14 streaming 
multiprocessors per GPU.}
We see that the time to run $n$ networks scales almost perfectly (it is 
essentially the time to run a single network) up to $n=14$. The time to run 
15--28 networks is then about twice the time to run a single network, the time 
to run 29--42 networks is about three times the time to run a single network, 
and so on.  The period of 14 concurrent networks in the steps reflects the 
availability of 14 streaming multiprocessors on the Kepler GPU 
microarchitecture.  (The step period was found to be 15 when tested on a GPU 
using the Fermi GPU microarchitecture with 15 streaming multiprocessors 
available.)  The slight rise in execution time on any given step presumably 
reflects a small increase in CPU overhead associated with launching increasing 
numbers of networks concurrently.

The results implied by \fig{stackedNetworkTiming} have large implications for 
simulations in a variety of scientific fields.  They demonstrate that not only 
can a single realistic network be run fast enough now to couple to fluid 
dynamics, but in fact {\em many} such networks can execute in a short enough 
time to make the simulation feasible.  The results presented in this paper are 
far from optimized (for example, we have only skimmed the surface of 
implementing efficient memory access), so we shall save detailed benchmarking 
for future papers.  However, based on the results presented here (see 
\fig{scaling_isotopes2} and \fig{stackedNetworkTiming} and the discussion associated with
Table \ref{table:netIntegrationTimes}), and those of Refs.\ 
\cite{guidJCP,guidAsy,guidQSS,guidPE}, we may estimate that even without 
further optimization 14 realistic 150-isotope networks can now be run  on a 
GPU in perhaps 10--20\% of the time that a standard implicit code running on a 
CPU can integrate {\em one} such network, and that $\sim$100 realistic 
150-isotope networks could be run in parallel entirely on a GPU in the same 
length of time that a standard implicit code could run one such network on a 
CPU.

Implicit algorithms are more complex than explicit algorithms because of the
required iteration and matrix inversions. It remains to be seen whether they
they can implement GPU versions having the same runtime qualities and
low-memory footprint of this explicit method.
In particular, problems with matrices of dimension a few hundred are memory 
bound  
because of a 
small number of computations relative to data required compared with problems 
with large matrices.  To date, ports of the implicit code Xnet \cite{raphcode} 
to combined CPU--GPU architectures run only part of the calculation on the GPU, 
and do not give appreciable increase in performance over implementations on the 
CPU alone \cite{harrisPrivate}.  Recent tests of an accelerated batched LU 
factorization of  $150 \times 150$ matrices using the MAGMA library 
hybrid CPU--GPU algorithm have demonstrated speed increases as large as a 
factor of $\sim 3-4$ \cite{dong2014}, but those methods remain to be fully 
implemented in Xnet \cite{harrisPrivate}.

The utility of these developments for realistic simulations is apparent.  In an
operator-split implementation of a zone-based fluid dynamics plus kinetics 
simulation, one could deploy the zones of the fluid dynamics on the CPUs of the 
compute nodes and the corresponding kinetic networks on the GPUs of the compute 
nodes (with one CPU per node allocating a small fraction of cycles to GPU 
management).  Then the GPU can execute independent kinetic networks in parallel 
for many of the fluid dynamics zones at once in a short enough time to make the 
calculation with realistic networks in many zones feasible.%
\footnote{
The parallel network launches on a given GPU are asynchronous.  To simplify this 
proof of principle the concurrent networks all were assigned the same 
temperature, density, and initial abundances, and so took the same number of 
integration steps. In a realistic application the concurrent networks in 
different zones typically would have the same reaction structure but perhaps 
different abundances and rates because of different conditions in each zone.  
Thus they may require varying numbers of integration steps and so might not 
return at the same time. The management of the network kernel launches can 
presumably be used to optimize workflow and load balancing in the coupled fluid 
dynamics and kinetic calculation.  We have not explored these issues yet. }
This goes far beyond the current limitations of being able to execute only 
a highly-restricted number of unrealistically small kinetic networks coupled to 
the fluid dynamics, and plausibly enables a variety of simulations with 
realistic kinetics that previously were  not accessible with available 
computing power.

\section{GPU Virtues and the Explicit Approach}

Let us revisit the four issues listed earlier in \S\ref{GPUacceleration} that 
require particular attention for the efficient utilization of GPU acceleration. 
As we have shown, the simplicity of the explicit approach has permitted us to 
address these issues in a highly expeditious manner.

(1)~Because the explicit method is compact, 
it is easy to fit entirely on the GPU so that no CPU--GPU communication is 
required except to launch the kernel and retrieve the results at the end of the 
calculation.  Furthermore, the required data transfer at the beginning and 
end is minimal (of order a few kB for the representative examples shown here).  
Thus, there is little CPU--GPU communication penalty, even for the launch of 
multiple concurrent networks.

(2)~The explicit algorithm is naturally highly parallel.  The calculation of 
rates from the rate library expressions is a very parallel operation 
since the rates are all independent and need be calculated only once for each 
network integration.%
\footnote{
In the operator splitting approximation that we are employing the temperature 
and density are held constant during the network integration, so the rates also 
remain constant during the network integration steps corresponding to one hydro 
timestep and need be calculated only once per hydro timestep.
}
The calculation of fluxes must be done at each network integration step (fluxes 
are products of rates and population variables; rates are constant but 
populations change with time in the integration), but that is also highly 
parallel since the fluxes are independent.  The least parallel part of applying 
the algebraically-stabilized explicit asymptotic algorithm is that the sums of 
fluxes populating and depopulating a given isotope (the $\fminus i$ and $\fplus 
i$ of \eq{eq1.1}) are required in order to compute the asymptotic update at each 
network timestep, and these summations are not intrinsically parallel.  However, 
we have used tree methods to implement them, which can in principle attain $\ln 
N$ execution times for the sum of $N$ fluxes. (We are far from that scaling now, 
presumably because our memory accesses are not yet optimized.)

(3)~Even with double precision it is possible to fit most of the important 
variables into fast shared memory because of the compactness of the algorithm. 
We have also demonstrated that single-precision integration is stable and 
sufficiently accurate for many applications,  which permits placing roughly 
twice as many variables in shared memory.  For example,  all relevant variables 
should fit in shared memory for the 150-isotope network in single precision. 

(4)~We have demonstrated that one can greatly increase thread occupancy by 
launching multiple networks concurrently. Thus the explicit method applied to 
multiple parallel networks scales to use all of the GPU streaming
multiprocessors.

\section{Future Work}

It is our intention to develop the present technology into an open-source, 
general-purpose code for solving  physically-realistic kinetic networks in a 
variety of disciplines.  Although the present results are a substantial step in 
that direction,  several important pieces remain to be implemented.

\begin{enumerate}
\item 
The present implementation uses only the explicit asymptotic algorithm.  This is 
adequate far from equilibrium but as systems approach equilibrium the asymptotic 
algorithm must be supplemented by a partial equilibrium algorithm 
to continue to scale \cite{guidPE}.  Implementation of the 
asymptotic plus partial equilibrium algorithm requires only some flux 
modifications and some additional bookkeeping, neither of which should have much 
impact on parallelism, so we do not anticipate major issues with implementing 
it.
\item
The present algorithm is not well optimized with respect to memory access
patterns. We anticipate that additional work on this issue
will lead to a substantial increase in speed. The most significant source of
concern in this area is coalescing memory accesses in order to prevent serialization
caused by uncoalesced memory.  It may be useful to utilize the GPU's texture memory,
which features a dedicated read-only cache optimized for spatial locality in a texture's
coordinate system rather than memory locality.
\item
We have shown that these calculations are stable and accurate with 
single-precision integration.  This should allow less memory usage and faster 
calculations than double-precision integration, but this has not yet been fully explored.
\item
The least parallel part of the algorithm is the summation of fluxes changing the 
population of each isotope in each integration step.  It is likely that this 
piece of the algorithm can be improved by restructuring, with a corresponding 
increase in speed.
\item
There are load-balancing issues associated with the flux summations that we have 
yet to address, occurring because the different isotopes in the network can 
require very different numbers of fluxes to be summed.  For example, in the 
larger networks used here protons, neutrons, and \isotope{4}{He}  each have 
hundreds of fluxes that change their populations and must be summed, whereas 
almost all of the other isotopes in the network have fewer than ten.  Presumably 
considerable optimization can be attained by a more load-balanced 
implementation 
of the flux summation algorithm.
\item
The explicit Quasi-Steady-State (QSS) algorithm is  similar to the asymptotic 
algorithm and may in some cases give better CPU integration performance than the 
asymptotic algorithm \cite{guidQSS}.  It will be of interest to see if 
replacement of the asymptotic algorithm with the QSS algorithm in the present 
code leads to improved GPU performance.
\item
This paper has dealt with using GPU accelerators to implement the kinetics 
integration.  We intend to port the current algorithm to systems with many-core 
accelerators, which we anticipate will also permit much more ambitious 
calculations than have been possible before.
\item 
This work did not explore the benefits of overlapping computation and
communication but it must be reviewed in the future.
\end{enumerate}

\noindent
Work is in progress on these improvements and we expect to report on 
them in future papers.

\section{Summary and Conclusions}

We have demonstrated that newly-developed explicit integration algorithms 
exhibiting many properties that are more desirable than those of implicit 
methods for large networks, coupled with multithreaded acceleration on GPUs, may 
permit orders of magnitude decreases in the runtimes to simulate large and 
realistic reaction kinetic networks that can be coupled to other physics 
solvers. Furthermore, we have demonstrated that for finding the solutions of 
many networks simultaneously the GPU far exceeds the performance of the CPU and 
its specific performance depends on the available number of streaming 
multiprocessors. These properties may make it possible to solve realistic 
coupled physics problems where the reaction network solve is the limiting 
factor. Finally, we described the future work that will be required to fully 
realize the performance benefits of using a GPU on these problems.

Readers interested in obtaining the source code should contact the corresponding
author directly.

\section{Acknowledgements}

We thank Taro Yamaguchi-Phillips, Raph Hix, Bronson Messer, Austin Harris, Tom 
Papatheodore, and Reuben Budiardja for discussions; and David Bernholdt,
Alexander J. McCaskey, and Phil Roth from Oak Ridge National Laboratory (ORNL)
for thoroughly reviewing the manuscript and suggesting many helpful corrections.

This work has been supported by the US Department of Energy, Office of
Nuclear Physics, and by the ORNL Undergraduate Research Participation Program,
which is sponsored by ORNL and administered jointly by ORNL and the Oak Ridge
Institute for Science and Education (ORISE). ORNL is managed by UT-Battelle, LLC,
for the US Department of Energy under contract no. DE-AC05-00OR22725. ORISE
is managed by Oak Ridge Associated Universities for the US Department of
Energy under contract no. DE-AC05-00OR22750.

\clearpage

\bibliographystyle{unsrt}

\end{document}